\documentclass[aps,floats,twocolumn,epsf,prl,showpacs,superscriptaddress]{revtex4-1}
\usepackage{graphicx}
\usepackage{epstopdf}
\usepackage{amsmath}
\usepackage{amssymb}
\usepackage[colorlinks=true,urlcolor=blue,citecolor=blue,linkcolor=blue,breaklinks=true]{hyperref}
\usepackage{color}
\usepackage[normalem]{ulem}
\usepackage[dvipsnames]{xcolor}

\usepackage{ulem}

\begin{document}

\title{
Superconducting landscape of the strongly correlated bilayer Hubbard model
}

\author{Motoharu Kitatani}
\affiliation{Graduate School of Engineering, Kyushu Institute of Technology, Kitakyushu 804-8550, Japan}
\affiliation{Department of Material Science, University of Hyogo, Ako, Hyogo 678-1297, Japan}

\author{Yusuke Nomura}
\affiliation{Institute for Materials Research (IMR), Tohoku University, Sendai 980-8577, Japan}
\affiliation{International Quantum Materials Center (Q$^2$), Advanced Institute for Materials Research (WPI-AIMR), Tohoku University, Sendai 980-8577, Japan}

\author{Shiro Sakai}
\affiliation{Physics Division, Sophia University, Chiyoda-ku, Tokyo 102-8554, Japan}

\author{Ryotaro Arita}
\affiliation{RIKEN Center for Emergent Matter Science (CEMS), Wako, Saitama 351-0198, Japan}
\affiliation{Department of Physics, The University of Tokyo, Hongo, Tokyo 113-0033, Japan}

\date{\today}

\begin{abstract}
Coupling two strongly correlated layers opens a distinct route to repulsion-driven superconductivity. Yet the same interlayer coupling makes the optimization of superconductivity highly nontrivial. Using the dynamical vertex approximation, we determine the superconducting landscape in the enlarged parameter space spanned by filling and interlayer hopping. We identify an optimal regime at intermediate interlayer hopping and moderate doping away from half filling, where the system exhibits high superconducting transition temperatures ($T_{\rm c}$) with an interband $s_{\pm}$-wave gap structure. Toward half filling, however, strong self-energy effects can suppress the superconducting tendency by reducing the low-energy spectral weight. We further show that a Fermi-surface distortion induced by longer-range hoppings can weaken this suppression and enhance $T_{\rm c}$. Our results establish how filling, interlayer hopping, and Fermi-surface geometry jointly optimize the superconducting regime in the strongly correlated bilayer Hubbard model.
\end{abstract}

\maketitle



\textit{Introduction.}---
Understanding how repulsive interactions can generate high-temperature superconductivity remains a central problem in condensed matter physics. In this context, coupling two strongly correlated layers opens a qualitatively distinct pairing route. In the minimal bilayer Hubbard model, the interlayer hopping ($t_\perp$) produces bonding and antibonding bands that can form two disconnected Fermi surfaces. Such a Fermi-surface geometry can stabilize an $s_{\pm}$-wave pairing state in which repulsive interband scattering acts as an effective pairing interaction and the superconducting gap changes sign between two bands \cite{Kuroki2002,Maier2011,Karakuzu2021}. This mechanism contrasts with the nodal $d$-wave pairing in the extensively studied single-layer Hubbard model \cite{Scalapino2012,LeBlanc2015,Schaefer2021}, where the sign change occurs within a single Fermi surface. By shifting the sign change from within a Fermi surface to between disconnected Fermi surfaces, the bilayer geometry can avoid nodes on the Fermi surface and may thereby support a higher transition temperature ($T_{\rm c}$) \cite{Kuroki2002}. This viewpoint has also gained renewed attention in connection with the recent discovery of superconductivity in bilayer nickelates such as La$_3$Ni$_2$O$_7$ \cite{Sun2023}, where bilayer Hubbard-type models have been considered as candidate low-energy frameworks \cite{Sakakibara2024,Maier2026}.

At the same time, the interlayer coupling makes the electronic structure nontrivial, particularly in the strongly correlated regime. At half filling, increasing $t_\perp$ tunes the competition between interaction-driven localization and interlayer hybridization, leading to a crossover from a Mott insulator to a band insulator \cite{Kancharla2007,Fujiwara2026arXiv}. Exploring superconductivity upon doping this strongly correlated regime is thus far from a simple extension of the single-layer problem. Interlayer coupling favors 
$s_{\pm}$-wave pairing between the bonding and antibonding bands. Yet too strong a coupling can push the antibonding band far from the Fermi level, weakening its contribution to pairing. This balance is further modified through both the real and imaginary parts of the self-energy. The real part shifts the bonding and antibonding bands relative to each other. Unlike a momentum-independent static self-energy shift in a single-band system, which can be absorbed into the chemical potential, such a relative shift can qualitatively reshape the Fermi surfaces. The imaginary part, on the other hand, suppresses the low-energy spectral weight. Moreover, both the real and imaginary parts have a dynamical structure that can change markedly with doping \cite{Nomura2025}. Since pressure can modify the interlayer hybridization in bilayer compounds, the same coupling that makes the problem both intriguing and nontrivial may also provide an experimentally accessible tuning parameter for correlated superconductivity.

Previous studies using quantum Monte Carlo (QMC), perturbative and variational approaches, and cluster extensions of dynamical mean-field theory (DMFT) \cite{Metzner1989,Georges1996} have provided important insights into the roles of interband scattering, interlayer hopping, and Fermi-surface geometry in bilayer superconductivity \cite{Scalapino2012,Bulut1992,Hetzel1994,Scalettar1994,dosSantos1995,Bouadim2008,Liechtenstein1995,Kuroki2002,Lanata2009,Mishra2016,Nakata2017,Matsumoto2020,Kato2020,Maier2011,Maier2019,Karakuzu2021,Liu2025,Nomura2025,Fujiwara2026arXiv}. Nevertheless, how these factors, together with correlation-induced Fermi-surface reconstruction and low-energy spectral suppression, jointly determine the optimal superconducting regime remains unclear. The central question is therefore where and how superconductivity is optimized over the filling--$t_\perp$ plane under correlation-driven reshaping of the electronic structure.

In this Letter, we apply the dynamical vertex approximation (D$\Gamma$A) \cite{Toschi2007,Katanin2009}, which incorporates long-range nonlocal fluctuations on top of local strong correlations, to the bilayer Hubbard model. We find that superconductivity is optimized at intermediate $t_\perp$ and moderate hole doping, where both bonding and antibonding Fermi surfaces participate in interband $s_{\pm}$-wave pairing. Toward half filling, however, the pairing interaction becomes stronger, while self-energy effects strongly suppress the low-energy spectral weight and reduce the superconducting tendency. We further show that, within the present framework, Fermi-surface distortion can mitigate the low-energy spectral suppression and enhance superconductivity. The resulting $T_{\rm c}$ can exceed that previously obtained for the single-layer model within D$\Gamma$A. These results highlight the bilayer Hubbard model as a rich and tunable platform, where control of interlayer coupling and Fermi-surface geometry can promote high-$T_{\rm c}$ superconductivity.

\textit{Model and Methods.}---
We consider the bilayer Hubbard model on a square lattice,
\begin{align}
    {\cal H}= -\sum_{i,j,l,\sigma} t_{ij} c^{\dagger}_{il\sigma}c_{jl\sigma}
    -&\sum_{i,\sigma} t_{\perp} \left(
    c^{\dagger}_{i1\sigma}c_{i2\sigma}+c^{\dagger}_{i2\sigma}c_{i1\sigma}
    \right) \nonumber \\ 
    &+U\sum_{i,l}n_{il\uparrow}n_{il\downarrow},
\end{align}
where $c^{\dagger}_{il\sigma} (c_{il\sigma})$ is the creation (annihilation) operator of an electron at site $i$ in layer $l\in\{1,2\}$ with spin $\sigma\in\{\uparrow,\downarrow \}$, $n_{il\sigma} \equiv c^{\dagger}_{il\sigma}c_{il\sigma}$, $t_{ij}$ is intralayer hopping (nearest-neighbor $t$, next-nearest-neighbor $t'$, and third-nearest-neighbor $t''$), $t_\perp$ is the interlayer hopping, and $U$ is the onsite interaction. It is convenient to represent the layer degree of freedom by a pseudo momentum $k_z \in \{0, \pi\}$, where we assign $k_z=0$ to the bonding and $k_z=\pi$ to the antibonding sector. Then, the non-interacting part of the Hamiltonian can be written as ${\cal H}_0 =\sum_{\mathbf{k},\sigma} \epsilon_{\mathbf{k}} c^{\dagger}_{\mathbf{k}\sigma}c_{\mathbf{k}\sigma}$ with the dispersion:

\begin{align}
    \epsilon_{\mathbf{k}} = -2t&(\cos{k_x}+\cos{k_y})-4t'\cos{k_x}\cos{k_y} \nonumber \\
    &-2t''(\cos{2k_x}+\cos{2k_y})-t_{\perp}\cos{k_z}.
\end{align}
The bilayer model naturally yields bonding and antibonding band splitting and can produce multiple Fermi surfaces, which is a key distinction from the single-layer model. In such systems with disconnected multiple Fermi surfaces, a sign-changing isotropic $s_{\pm}$-wave superconducting state between different Fermi surfaces has been discussed \cite{Kuroki2002,Maier2011}.


For analyzing this model, we employ the D$\Gamma$A with the $\lambda$-correction \cite{Toschi2007,Katanin2009}, a diagrammatic extension of DMFT \cite{Metzner1989,Georges1996} that incorporates long-range nonlocal fluctuations on top of the strong-correlation effects captured by DMFT, where we use \texttt{w2dynamics} \cite{Wallerberger2019} as an impurity solver. To analyze superconductivity, we evaluate the leading eigenvalue $\lambda_{\mathrm{SC}}$ of the linearized gap equation in the particle-particle channel. For further technical details, we refer the reader to the reviews in Refs.~\cite{Rohringer2018,Kitatani2022}. This method has been successfully applied to Hubbard-model studies of cuprate and infinite-layer nickelate superconductivity and is suitable for the present purpose, because it allows us to treat antiferromagnetic-fluctuation-mediated pairing and pseudogap-like spectral renormalization within the same framework \cite{Kitatani2019,Kitatani2020}. 
Here, we apply this framework to the bilayer Hubbard model. In the actual calculations, we treat the bilayer degrees of freedom in the $k_z=0$ and $k_z=\pi$ basis, as introduced above, which effectively corresponds to solving a Hubbard model with $n_k \times n_k \times 2$ discrete momentum degrees of freedom. Thus, the local problem is treated within single-site DMFT, while nonlocal correlations, including in-plane and interlayer fluctuations, are incorporated through the diagrammatic extension of D$\Gamma$A.

\begin{figure}[tbp]
        \centering
        \includegraphics[width=\linewidth,angle=0]{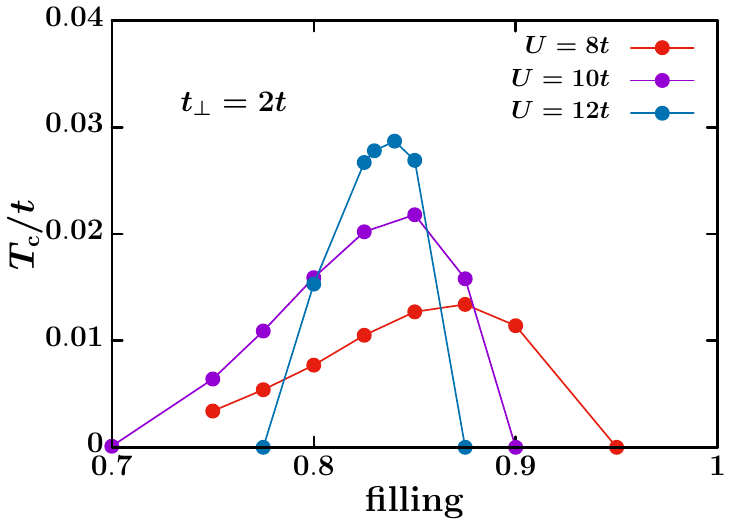}
        \caption{Superconducting phase diagram ($T_{\rm c}$ vs filling) of the bilayer Hubbard model for $U=8t, 10t, 12t$, with $t'=t''=0$ and $t_\perp =2t$.}
        \label{fig:phasediagram}
\end{figure}


\textit{Results.}---
We start by presenting the superconducting phase diagram for $U/t=8,10,12$ and $t'=t''= 0$, as shown in Fig.~\ref{fig:phasediagram}. Here, the superconducting $T_{\rm c}$ is determined as the temperature at which
the leading eigenvalue $\lambda_{\mathrm{SC}}$ of the linearized gap equation reaches unity \cite{extrapolation}. We obtain a dome-shaped behavior of $T_{\rm c}$ around $n=0.85$. As the interaction $U$ increases, the peak $T_{\rm c}$ increases, while the dome becomes narrower. Overall, the maximum $T_{\rm c}$ in the bilayer system 
is higher than
that in the single-layer model \cite{Kitatani2023}, consistent with previous studies using other methods \cite{Kuroki2002}. On the other hand, we find a strong suppression of $T_{\rm c}$ upon approaching half filling. This suppression is a key feature of the strongly correlated regime and motivates the following analysis of the low-energy spectral weight and self-energy.

\begin{figure}[bp]
        \centering
        \includegraphics[width=\linewidth,angle=0]{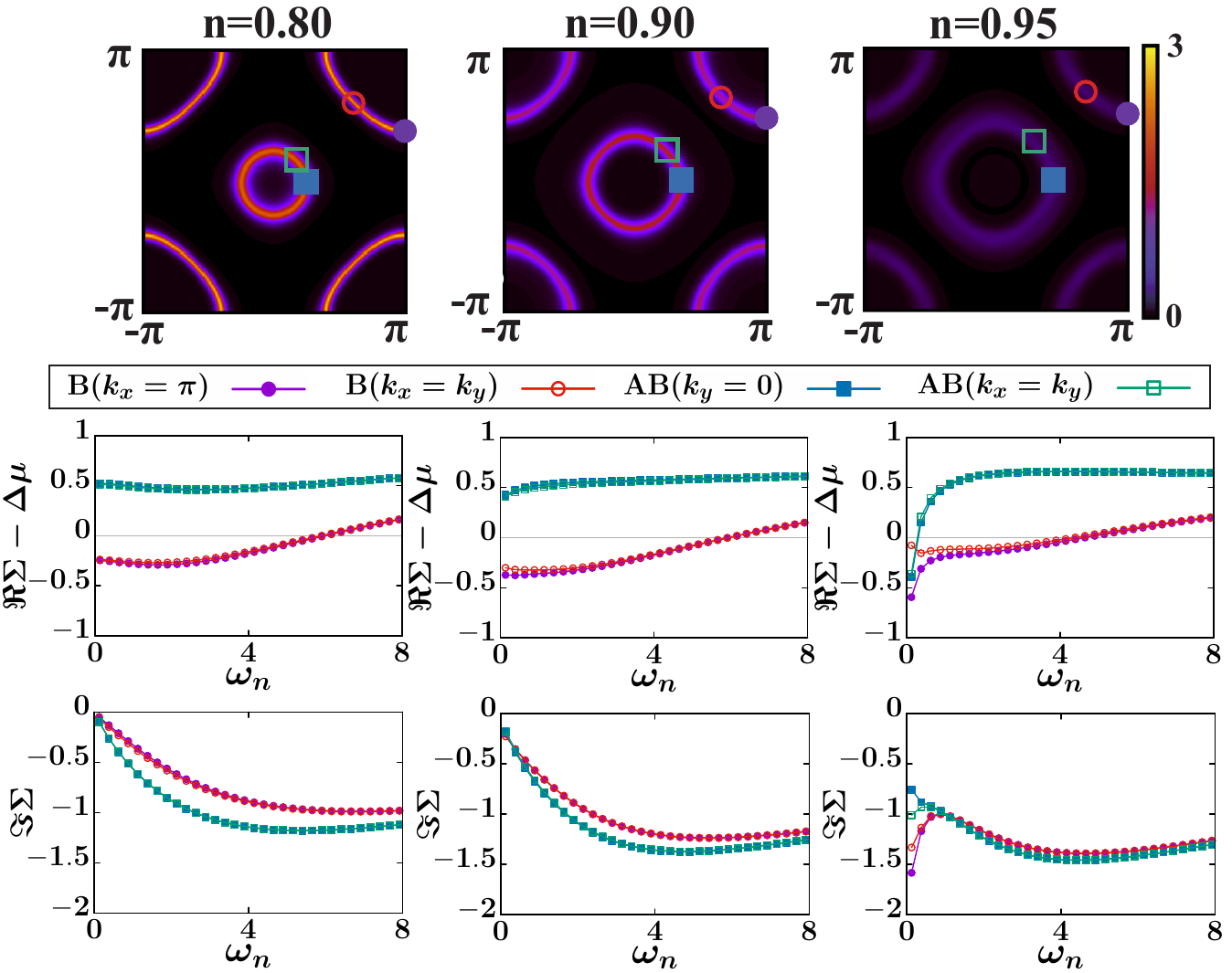}
        \caption{Filling dependence ($n=0.80, 0.90, 0.95$) of the spectrum ($-\beta G(\tau=\beta/2)/\pi$) and the self-energy for $U=8t, t_\perp=2t$, $t'=t''=0$, and $T=0.04t$. Here, $\Delta \mu$ denotes the difference in chemical potential between interacting ($U=8t$) and non-interacting ($U=0$) systems.}
        \label{fig:spectrum_fillingdep}
\end{figure}


In Fig.~\ref{fig:spectrum_fillingdep}, the top row shows the momentum-space maps of the low-energy spectrum (summed over the bonding and antibonding bands) for $n=0.80, 0.90$, and $0.95$, while the middle and bottom rows present the Matsubara-frequency [$\omega_n=(2n+1)\pi T$ with integer $n$ and temperature $T=0.04t$] dependence of the self-energy. The self-energy is evaluated at four representative momenta, indicated by open/filled circles/squares in the spectral plots (two points on each Fermi surface for the bonding and antibonding bands, respectively, taken along the $k_x=k_y$ direction and along a momentum cut with $k_x=\pi$ ($k_y=0$) fixed for the bonding (antibonding) Fermi surface).
The middle row shows the real part of the self-energy in the bonding and antibonding sectors, suggesting that correlations renormalize the effective bonding-antibonding splitting.
For $n=0.80$ and $n=0.90$, the low-frequency $\Re \Sigma-\Delta \mu$ is negative in the bonding sector and positive in the antibonding sector, which increases the effective bonding-antibonding splitting. At these fillings, $\Im \Sigma$ exhibits Fermi-liquid-like behavior, with its magnitude decreasing toward low frequencies, consistent with large spectral weight near the Fermi level. In contrast, at $n=0.95$, $\Re \Sigma$ develops a pronounced dynamical structure at low frequencies.
While this effect pulls the antibonding band back toward the Fermi level, the spectral weight is significantly suppressed due to the strong scattering. This suppression is consistent with an enhancement of $|\Im \Sigma|$ at low frequencies in the bottom row of Fig.~\ref{fig:spectrum_fillingdep}.

Indeed, $T_{\rm c}$ is strongly reduced toward half filling, and this reduction coincides with the regime where the low-energy spectral weight suppression becomes prominent. In particular, at $n=0.95$, the spectral weight is strongly suppressed, and correspondingly, $T_{\rm c}$ is more strongly suppressed near half filling in the present D$\Gamma$A results than in previous fluctuation exchange (FLEX) and cluster DMFT studies \cite{Kuroki2002,Matsumoto2020,Nomura2025}. A similar tendency toward a more pronounced suppression near half filling in D$\Gamma$A than in cluster calculations has also been seen in the single-layer Hubbard model \cite{Chen2015,Kitatani2023}.

\begin{figure}[tbp]
        \centering
        \includegraphics[width=\linewidth,angle=0]{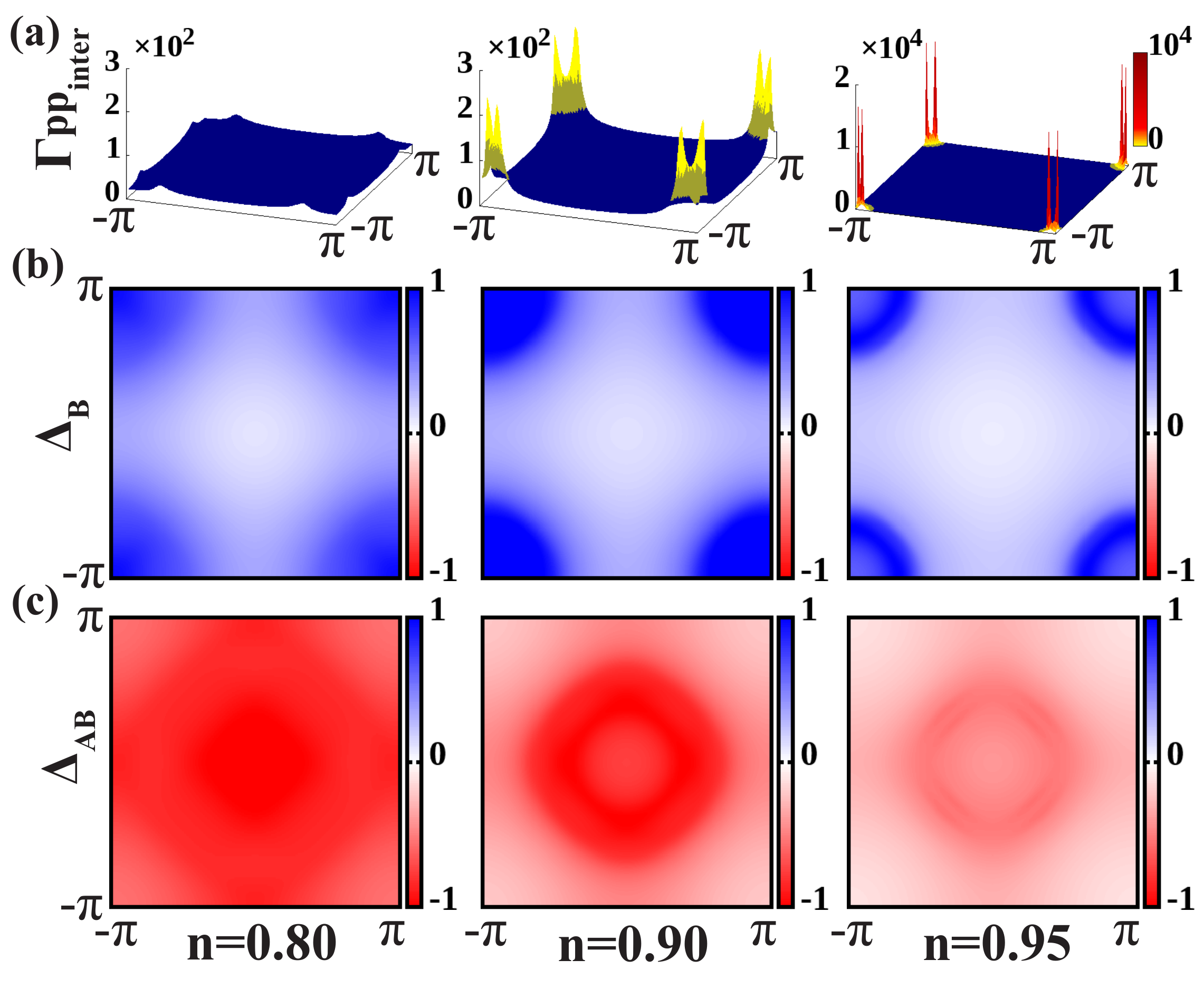}
        \caption{(a) Pairing interaction for interband scattering $\Gamma_{{\rm pp}_{\rm inter}}$, evaluated at the lowest fermionic Matsubara frequencies and zero bosonic transfer frequency in the spin-fluctuation channel, for $U=8t, t_\perp=2t$, $t'=t''=0$, and $T=0.04t$. Each column corresponds to the filling $n$ = 0.80, 0.90, and 0.95. (b,c) Bonding and antibonding components ($\Delta_{\rm B}$ and $\Delta_{\rm AB}$, respectively) of the eigenvector corresponding to the leading eigenvalue ($\lambda_{\mathrm{SC}}=0.56,0.62$, and $0.26$, from left to right) of the linearized Eliashberg equation for the same parameters as in (a). Eigenvectors are shown at the lowest positive Matsubara frequency, and normalized by their maximum absolute value for each filling.}
        \label{fig:gap_fillingdep}
\end{figure}


Next, we analyze the pairing interaction and the gap symmetry shown in Fig.~\ref{fig:gap_fillingdep}. We first discuss the filling dependence of the pairing interaction in Fig.~\ref{fig:gap_fillingdep}(a). In the bilayer system, the interband contribution, i.e., scattering between the bonding and antibonding bands, is dominant, and we therefore show only this part. Its momentum dependence exhibits a pronounced peak around $(\pi, \pi)$, indicating that antiferromagnetic spin fluctuations are strongly developed.

Then, we examine the gap structure, represented by the eigenvector associated with the leading eigenvalue of the linearized gap equation, in Fig.~\ref{fig:gap_fillingdep}(b,c). We obtain an $s_\pm$-wave symmetry that preserves the fourfold (C$_4$) symmetry within each layer but with a sign change of the gap between different Fermi surfaces. In the overdoped regime, the gap exhibits a relatively smooth momentum dependence, whereas closer to half filling, it shows a more structured momentum dependence with a sharper peak on the Fermi surface.

While the pairing interaction itself tends to increase toward half filling, the low-energy spectral weight is simultaneously reduced by pseudogap-like self-energy effects. As a result, $T_{\rm c}$ is strongly suppressed near half filling, and a dome-shaped structure emerges in a manner similar to that discussed for the single-layer Hubbard model.

\begin{figure}[tbp]
        \centering
        \includegraphics[width=\linewidth,angle=0]{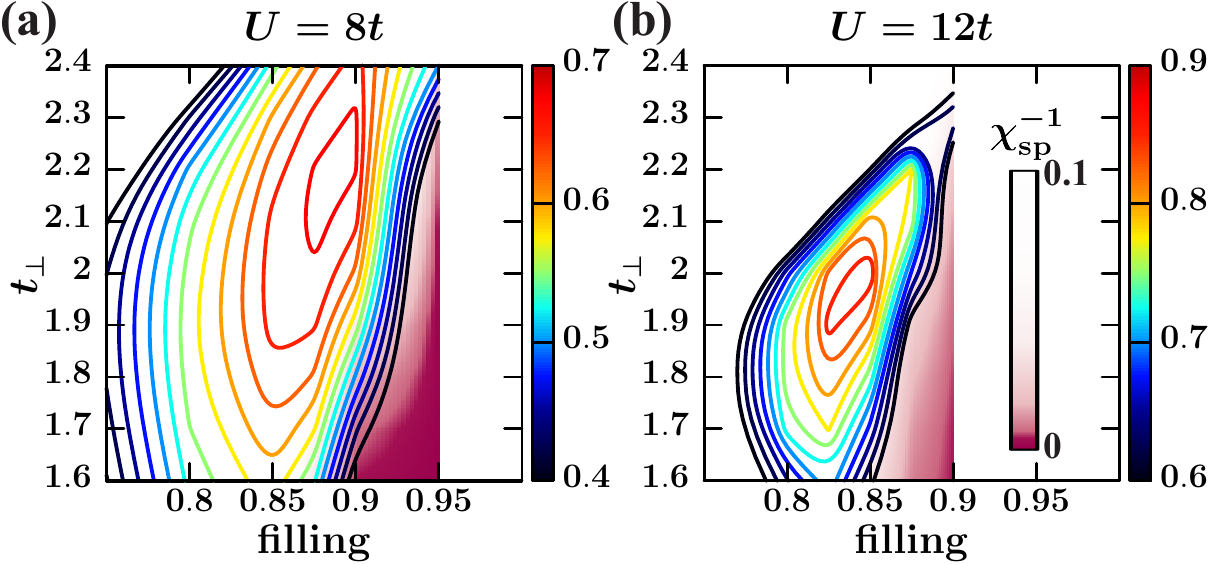}
        \caption{Filling and $t_\perp$ dependence of the superconducting eigenvalue $\lambda_{\mathrm{SC}}$ (contours) and antiferromagnetic susceptibility $\chi_{\rm sp}(\mathbf{Q}_{\rm max},\omega=0)$ for (a) $U=8t$ and (b) $U=12t$ with $t'=t''=0$ and $T=0.04t$, where $\mathbf{Q}_{\rm max}$ is the wave vector at which the static spin susceptibility is maximal.}
        \label{fig:lambda_filling-tperpdep}
\end{figure}


Next, we show in Fig.~\ref{fig:lambda_filling-tperpdep} the contour plots of the leading superconducting eigenvalues $\lambda_{\mathrm{SC}}$ against $t_\perp$ and filling $n$ for a fixed temperature $T=0.04t$ and interactions (a) $U=8t$ and (b) $U=12t$ \cite{contourplot}. As in the phase diagram (Fig.~\ref{fig:phasediagram}), the peak value is higher for $U=12t$, while the parameter region where superconductivity is enhanced becomes narrower. On the other hand, the overall contour shape is nearly the same for the two values of $U$, indicating that the condition for the optimal pairing does not change qualitatively with $U$. These features can be understood by considering the filling and $t_\perp$ dependences separately, including the roles of the imaginary and real parts of the self-energy. 

Along the filling direction, the region of enhanced $\lambda_{\rm SC}$ is bounded toward half filling by the spectral weight suppression associated with the imaginary part of the self-energy, as illustrated in Fig.~\ref{fig:spectrum_fillingdep}. The background shading in Fig.~\ref{fig:lambda_filling-tperpdep} further shows that this suppression is accompanied by a strong enhancement of antiferromagnetic fluctuations. 

Along the $t_\perp$ direction, superconductivity is not sufficiently enhanced when $t_\perp$ is too small, whereas for too large $t_\perp$, the antibonding band no longer crosses the Fermi level. The real part of the self-energy also increases the effective bonding-antibonding splitting and can thereby shift the optimal $t_\perp$. For $U=12t$, the optimal $t_\perp$ lies in the range $1.9t$--$2.0t$, slightly below that for $U=8t$, reflecting a stronger real-part renormalization at larger $U$. Importantly, the maximum of $\lambda_{\mathrm{SC}}$ is found in the regime where both the bonding and antibonding Fermi surfaces are clearly present. Therefore, the optimization of superconductivity in the bilayer system requires an intermediate $t_\perp$ where both bands actively participate. This tendency for the peak to occur in the region where two Fermi surfaces coexist is consistent with previous dynamical cluster approximation (DCA) results \cite{Karakuzu2021,Liu2025}, though it may not align with the optimal region identified by FLEX.


We compare the D$\Gamma$A and FLEX results in more detail in the Supplemental Material \cite{SM}, including a direct comparison between them at $U=4t$ (Fig.~S.1) and the corresponding FLEX spectral and superconducting properties at $U=8t$ (Figs.~S.2--S.4). Overall, FLEX tends to yield a stronger pairing interaction, except for the pseudogap-like region, while its self-energy strongly affects the low-energy electronic structure. By comparing the effects of the pairing interaction and the self-energy entering the linearized gap equation, we also disentangle the contributions of the pairing interaction itself and the self-energy-renormalized Green function to the superconducting eigenvalue (Fig.~S.5). The FLEX results are strongly affected by Fermi-surface reshaping due to the real part of the self-energy, which increases the bonding-antibonding splitting and can lead to the disappearance of the electron pocket. In contrast, the suppression near half filling in D$\Gamma$A appears to be more closely associated with the depletion of low-energy spectral weight due to strong scattering. These results suggest that strong-correlation-induced spectral suppression should be considered alongside the interband pairing mechanism in the strongly correlated bilayer regime.

\begin{figure}[tbp]
        \centering
        \includegraphics[width=\linewidth,angle=0]{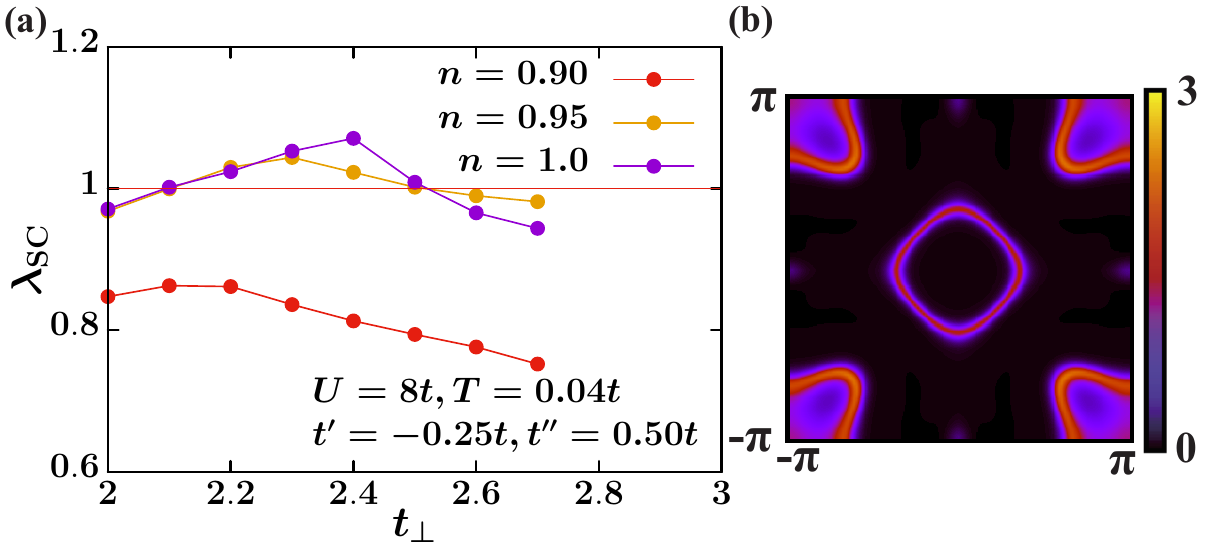}
        \caption{(a) Superconducting eigenvalue $\lambda_{\mathrm{SC}}$ as a function of $t_\perp$ for $n=0.90, 0.95, 1.0$. (b) Spectrum with warped Fermi surface ($-\beta G(\tau=\beta/2)/\pi$) for $n=0.95$ and $t_\perp=2.5t$.  Parameters are $U=8t, t^{\prime}=-0.25t$, $t^{\prime\prime}=0.50t$, and $T=0.04t$.
        }
        \label{fig:tp-0.25tpp0.50}
\end{figure}


Finally, we discuss how the superconducting tendency can be further enhanced. Given that our calculations exhibit a strong suppression of the low-energy spectrum driven by strong antiferromagnetic fluctuations, a natural strategy is to weaken the Fermi-surface nesting. Indeed, in the single-layer Hubbard model, it has been shown that introducing relatively large $t'$ and $t''$ leads to a peak in the D$\Gamma$A transition temperature near half filling \cite{Kitatani2023}. Motivated by this, in Fig.~\ref{fig:tp-0.25tpp0.50} we show the results for $n=0.90, 0.95, 1.0$ with large $t'$ and $t''$. Note that since the Fermi surface in the bilayer model is located near $\mathbf{k}=(0,0)$ and $\mathbf{k}=(\pi,\pi)$, the effect of $t'$ and $t''$ on the Fermi-surface shape is relatively small, and we therefore choose sufficiently large $t'$ and $t''$ to induce a clear Fermi-surface distortion. 

As a result, we obtain $T_{\rm c} \approx 0.04t$ at $U=8t$, which is higher than the value in Fig.~1, where $T_{\rm c}$ is already higher than that of the single-layer system. In fact, the spectrum in Fig.~\ref{fig:tp-0.25tpp0.50}(b) indicates that no gap-like behavior appears even near half filling. This result suggests that weakening this spectral weight suppression through Fermi-surface control can enhance superconductivity in the strongly correlated bilayer regime. Together with previous studies \cite{Kuroki2002,Maier2011,Karakuzu2021}, this finding supports the view that bilayer systems have a strong potential to achieve higher $T_{\rm c}$ than their single-layer counterparts. In addition, the eigenvector exhibits no strong momentum dependence, indicating that the interlayer pairing, rather than detailed in-plane structures, plays a dominant role.


\textit{Conclusion.}---
Using D$\Gamma$A, we have mapped the superconducting tendency in an enlarged parameter space of the strongly correlated bilayer Hubbard model. Here, interlayer coupling simultaneously controls the interband pairing channel and the Fermi-surface structure, and the interplay of these factors with self-energy effects makes the optimization of superconductivity nontrivial. In the absence of longer-range hopping, superconductivity is optimized at intermediate $t_\perp$ and moderate doping away from half filling. In this regime, both bonding and antibonding Fermi surfaces participate in interband $s_{\pm}$-wave pairing, and the system can exhibit high superconducting transition temperatures. Along the $t_\perp$ direction, the real part of the self-energy further increases the effective bonding-antibonding splitting. Toward half filling, however, the pairing interaction becomes stronger, while the imaginary part of the self-energy strongly suppresses the low-energy spectral weight, thereby limiting superconductivity on the half-filled side.

We further showed that Fermi-surface distortion induced by longer-range hopping can weaken the low-energy spectral weight suppression and enhance superconductivity, which suggests a possible route toward higher $T_{\rm c}$. Together, our results show how filling, interlayer hopping, and Fermi-surface geometry jointly determine the optimal superconducting regime, establishing interlayer hopping as an active control parameter in the strongly correlated bilayer Hubbard model.

An important next step is to treat the interlayer correlations more accurately. One possible direction is a diagrammatic extension starting from a $1\times1\times2$ cluster DMFT calculation. In the bonding-antibonding basis, this framework becomes equivalent to a DMFT extension for multi-orbital models and should therefore be feasible \cite{Kaufmann2021}. It will also be important to connect the strong sensitivity to Fermi-surface geometry found here to material-dependent band structures and trends in real compounds.

\textit{Acknowledgments.}---
This work was supported by JSPS KAKENHI (Grant Numbers JP24K17014, JP25K00961, JP26K00650,
JP25H01246, JP25H01252, JP24H00190,
JP23H04869, JP25H01506, JP26K00646, JP26K21723),
JST K-Program (Grant Number: JPMJKP25Z7), RIKEN TRIP initiative (RIKEN Quantum, Advanced General Intelligence for Science Program, Many-body Electron Systems)
and JST-ASPIRE (Grant Number: JPMJAP2512).
The computations were partially carried out using the computer resources offered under the category of General Projects by the Research Institute for Information Technology, Kyushu University, and those provided by the Supercomputer Center, the Institute for Solid State Physics, the University of Tokyo.

\bibliography{main}
\end{document}


\title{Supplemental Material for ``Superconducting landscape of the strongly correlated bilayer Hubbard model"
}

\author{Motoharu Kitatani}
\affiliation{Graduate School of Engineering, Kyushu Institute of Technology, Kitakyushu 804-8550, Japan}
\affiliation{Department of Material Science, University of Hyogo, Ako, Hyogo 678-1297, Japan}

\author{Yusuke Nomura}
\affiliation{Institute for Materials Research (IMR), Tohoku University, Sendai 980-8577, Japan}
\affiliation{International Quantum Materials Center (Q$^2$), Advanced Institute for Materials Research (WPI-AIMR), Tohoku University, Sendai 980-8577, Japan}

\author{Shiro Sakai}
\affiliation{Physics Division, Sophia University, Chiyoda-ku, Tokyo 102-8554, Japan}

\author{Ryotaro Arita}
\affiliation{RIKEN Center for Emergent Matter Science (CEMS), Wako, Saitama 351-0198, Japan}
\affiliation{Department of Physics, The University of Tokyo, Hongo, Tokyo 113-0033, Japan}

\date{\today}

\maketitle

\section{Comparison with FLEX results}
Here, we compare the D$\Gamma$A results with those obtained by fluctuation exchange (FLEX) approximation \cite{Bickers1989}. We first present a comparison in the intermediate-coupling regime, $U=4t$, where strong correlation effects are still moderate. Figure S.1 shows the spectral weight, $-\beta G(\tau=\beta/2)/\pi$, and the interband pairing interaction for $t_\perp =2t$ and $\beta t=10$. Although the D$\Gamma$A and FLEX show a qualitatively similar trend, FLEX provides a somewhat broader spectral structure and larger pairing interaction.

\begin{figure}[htbp]
        \centering
        \includegraphics[width=\linewidth,angle=0]{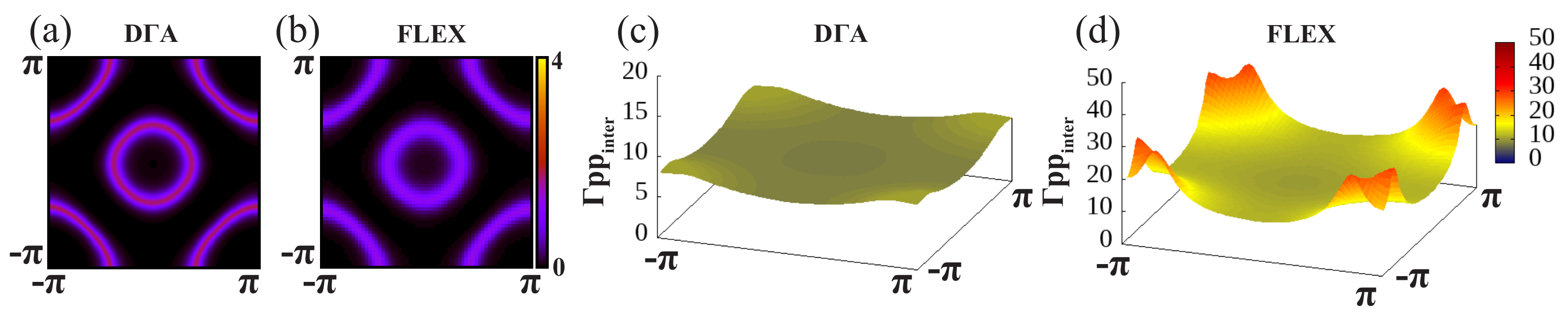}
        \caption{Comparison between D$\Gamma$A and FLEX for $U=4t, t_\perp=2t, n=0.85$, and $\beta t=10$. Left panels (a,b) show the spectral weight $-\beta G(\tau=\beta/2)/\pi$ obtained by D$\Gamma$A and FLEX, respectively. Right panels (c,d) show the interband pairing interaction.}
\end{figure}

\begin{figure}[htbp]
        \centering
        \includegraphics[width=\linewidth,angle=0]{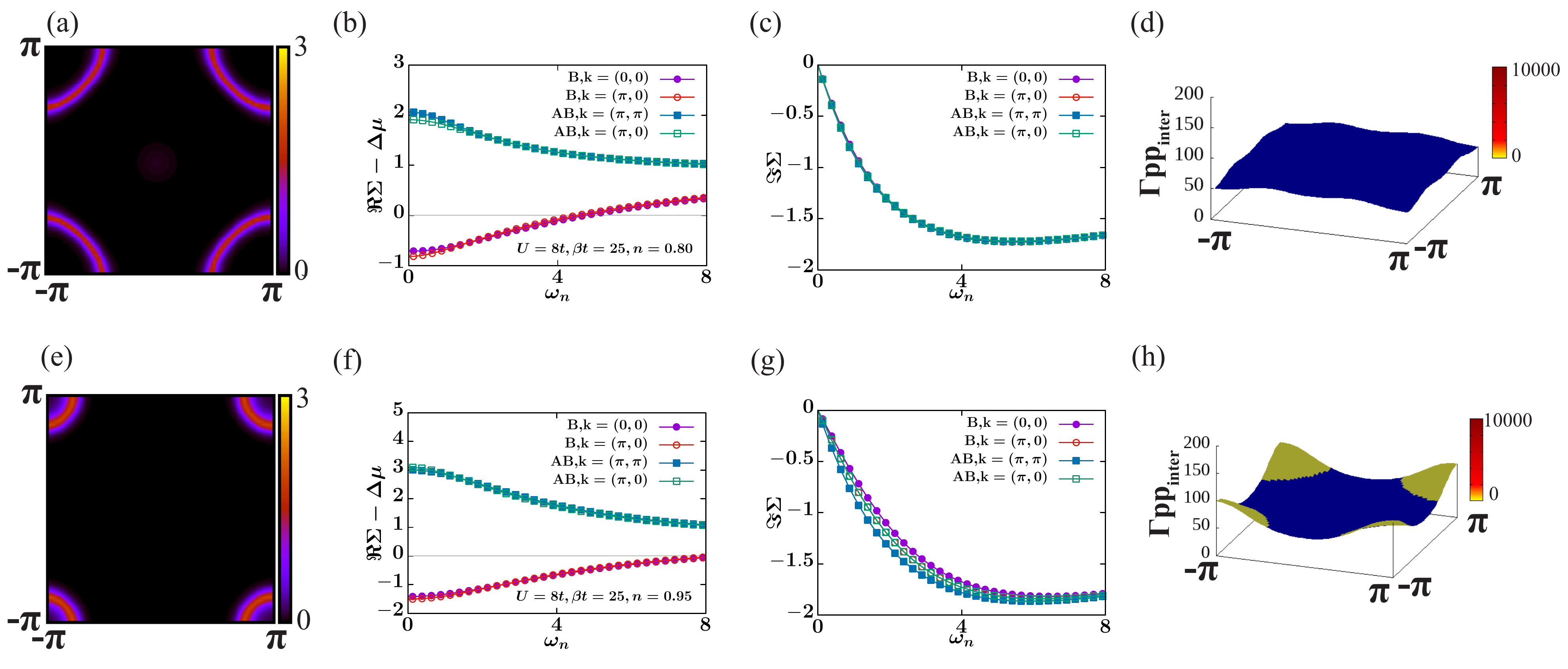}
        \caption{(a,e) FLEX spectral weight $-\beta G(\tau=\beta/2)/\pi$, (b,f) real part of the self-energy relative to $\Delta \mu$ ($\Delta \mu$ denotes the difference in chemical potential between interacting and non-interacting systems), (c,g) imaginary part of the self-energy, and (d,h) interband pairing interaction for $T=0.04t, t_\perp=2t,  t'=t''=0$ and $U=8t$. Panels (a--d) show the result at $n=0.80$, while panels (e--h) show those at $n=0.95$.}
        \label{fig:resultU8}
\end{figure}

Next, we show the FLEX spectrum, the real and imaginary parts of the self-energy, and interband pairing interaction at $U=8t$ in Fig.~\ref{fig:resultU8}, corresponding to Figs.~2 and 3 in the main text, using the same model parameters. We note that FLEX enhances the bonding-antibonding splitting, leading to the disappearance of the electron pocket. 
This tendency can also be seen directly in the real part of the FLEX self-energy. As shown in Fig.~\ref{fig:resultU8}(b,f), $\Re\Sigma$ produces different low-frequency shifts for the bonding
and antibonding components. Thus, the relative shift increases the effective bonding-antibonding splitting, providing a microscopic origin of the strong Fermi-surface reconstruction in FLEX.
Correspondingly, the pairing interaction exhibits a more momentum-independent structure [Fig.~\ref{fig:resultU8}(d,h)].

\begin{figure}[htbp]
        \centering
        \includegraphics[width=0.8\linewidth,angle=0]{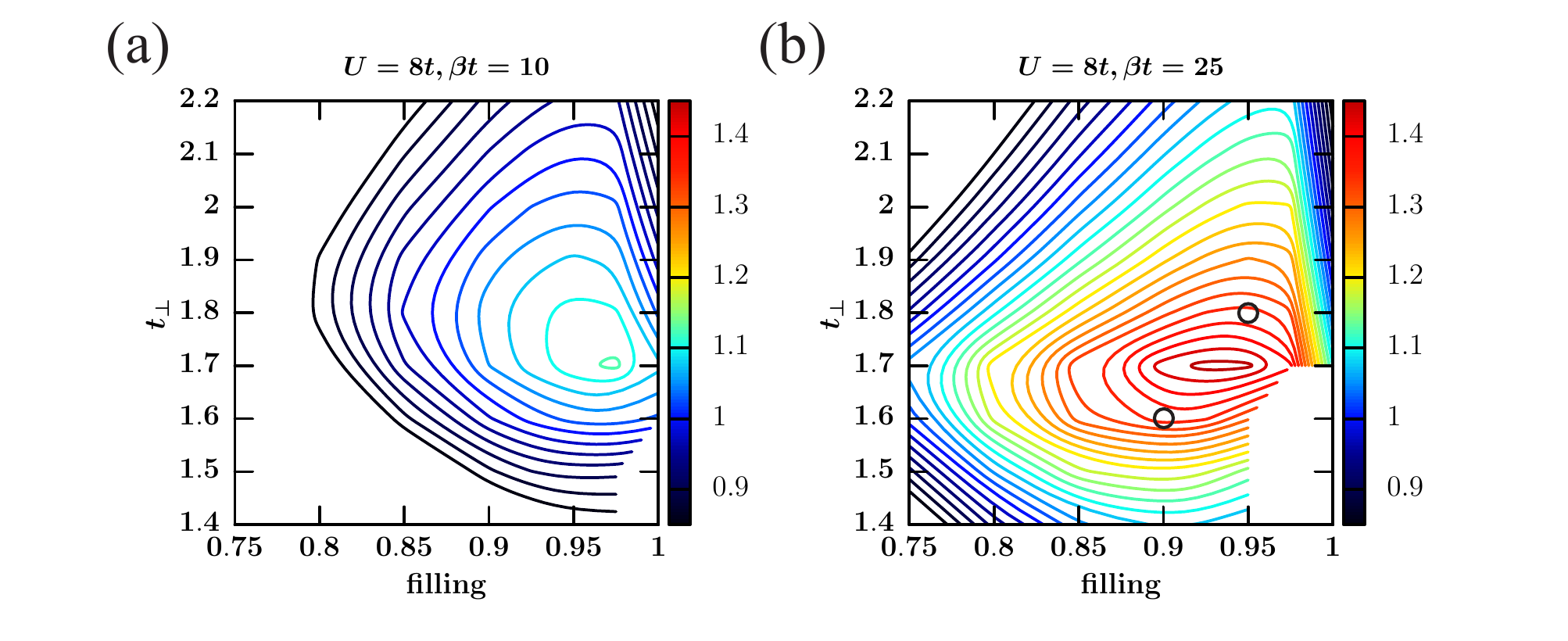}
        \caption{Contour plots of the FLEX superconducting eigenvalue $\lambda_{\mathrm{SC}}$ in the filling--$t_\perp$ plane for $U=8t$ at (a) $\beta t =10$ and (b) $\beta t =25$, respectively. The open circles in panel (b) indicate the two parameter sets compared in Fig.~\ref{fig:FLEXtempdep}.}
        \label{fig:FLEXcontour}
\end{figure}

We also show the contour plots of FLEX superconducting eigenvalues as functions of the filling and $t_\perp$ in Fig.~\ref{fig:FLEXcontour}. Around the superconducting transition temperature $T_{\rm c}$, the $t_\perp$ dependence is rather weak. This behavior is consistent with previous FLEX studies, which suggested that the appearance of the electron pocket does not lead to a sharp enhancement of the superconducting tendency \cite{Matsumoto2020}. Toward lower temperatures, the region with enhanced $\lambda_{\mathrm{SC}}$ extends toward the smaller $t_\perp$.

\begin{figure}[htbp]
        \centering
        \includegraphics[width=0.8\linewidth,angle=0]{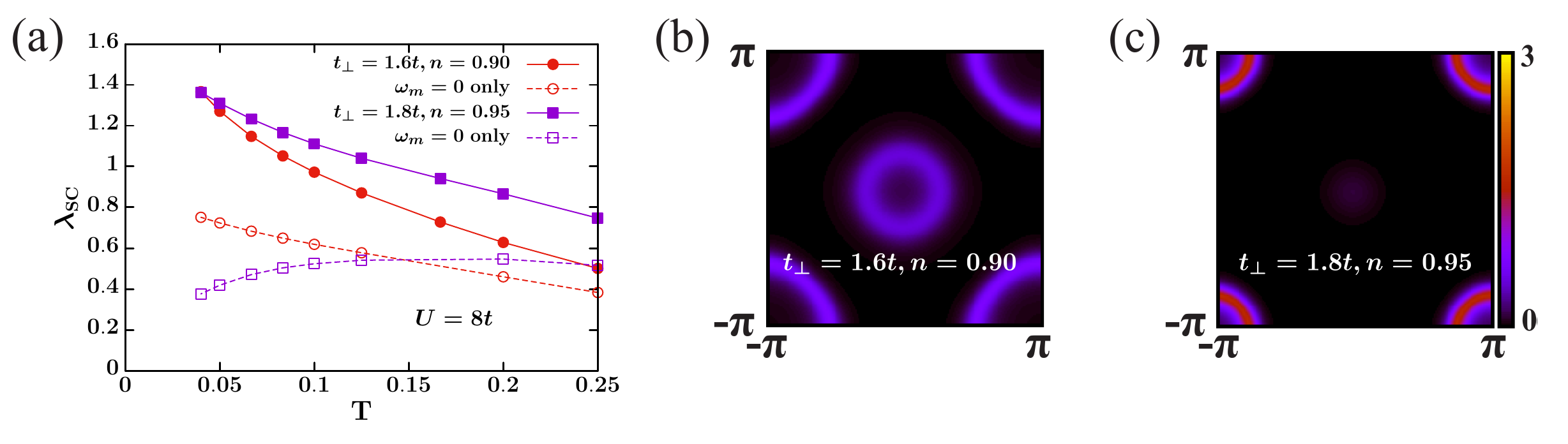}
        \caption{(a) Superconducting eigenvalue $(\lambda_{\mathrm{SC}})$ as a function of temperature for $U=8t$ at two parameter sets: $t_\perp=1.6t, n=0.90$ and $t_\perp=1.8t, n=0.95$. Filled symbols show the original FLEX results, whereas open symbols show the results obtained by retaining only the zero-bosonic-frequency component of the fluctuation-mediated part of the pairing interaction in the linearized Eliashberg equation, with the Green function kept unchanged in the latter calculations. (b,c) Low-energy spectral weight at $T=0.04t$ for $t_\perp=1.6t, n=0.90$ and $t_\perp=1.8t, n=0.95$, respectively.}
        \label{fig:FLEXtempdep}
\end{figure}

To examine this trend more directly, we compare the temperature dependence of $\lambda_{\mathrm{SC}}$ for two representative parameter sets in Fig.~\ref{fig:FLEXtempdep}. For $t_\perp=1.6t$ and $n=0.90$, where the two low-energy Fermi-surface features are clearly visible, $\lambda_{\mathrm{SC}}$ shows a stronger increase upon cooling and becomes comparable at low temperatures to that for
$t_\perp=1.8t$ and $n=0.95$. This behavior is consistent with the contour plots, which show that the enhanced-$\lambda_{\mathrm{SC}}$ region extends toward smaller $t_\perp$, especially in the moderately doped regime.

We also solve the linearized Eliashberg equation by retaining only the zero-bosonic-frequency component of the fluctuation-mediated part of the pairing interaction, while keeping the Green function unchanged. The resulting $\lambda_{\mathrm{SC}}$ is reduced for both parameter sets. For $t_\perp=1.6t$ and $n=0.90$, $\lambda_{\mathrm{SC}}$ still increases upon cooling, but less strongly than in the full calculation. In contrast, for $t_\perp=1.8t$ and $n=0.95$, the zero-frequency-only result decreases toward the lowest temperatures, whereas the full result continues to increase. These results indicate that the finite-frequency components of the pairing interaction make an important contribution to the low-temperature enhancement of $\lambda_{\mathrm{SC}}$ \cite{Matsumoto2020}, particularly for $t_\perp=1.8t$ and $n=0.95$.

We note that this comparison is not intended to provide quantitative trends of FLEX $T_{\rm c}$ from $\lambda_{\mathrm{SC}}\simeq 1$. Since D$\Gamma$A gives a lower transition temperature and a stronger spectral weight suppression near half filling, the superconducting enhancement found in D$\Gamma$A may be more naturally associated with a regime closer to Fig.~\ref{fig:FLEXtempdep}(b), where the two low-energy spectral features remain discernible.

\begin{figure}[htbp]
        \centering
        \includegraphics[width=0.8\linewidth,angle=0]{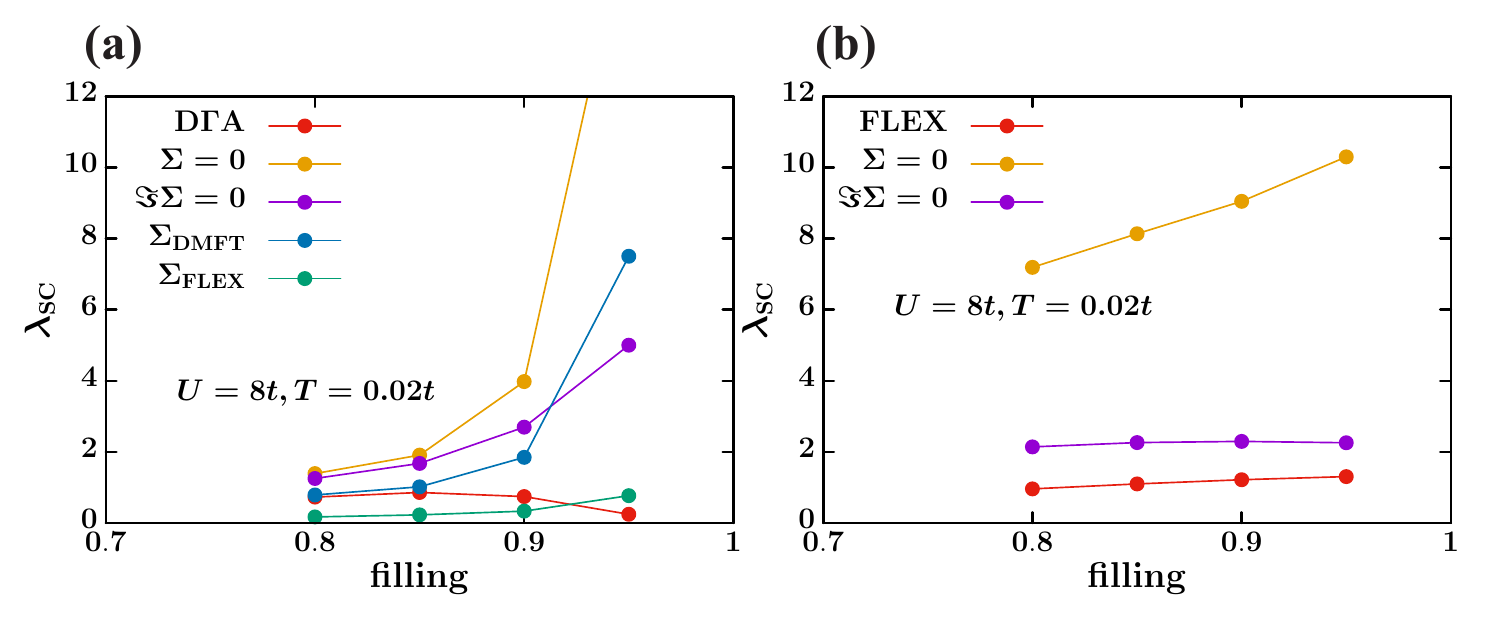}
        \caption{Comparison of the eigenvalue $\lambda_{\mathrm{SC}}$ between the original calculations [(a) D$\Gamma$A, (b) FLEX] and the calculations with the manually modified self-energies in the linearized gap equation, for $U=8t$, $T=0.02t$, and $t_\perp=2.0t$.}
        \label{fig:modified-self-energy}
\end{figure}

To study the origin of the differences in the superconducting tendency between D$\Gamma$A and FLEX in more detail, we disentangle the roles of the self-energy and the pairing interaction in determining the superconducting eigenvalue $\lambda_{\mathrm{SC}}$. We keep the pairing interaction obtained within D$\Gamma$A fixed in Fig.~\ref{fig:modified-self-energy}(a) and that obtained within FLEX fixed in Fig.~\ref{fig:modified-self-energy}(b), while varying only the self-energy entering the Green function in the linearized Eliashberg equation, and show the resulting $\lambda_{\mathrm{SC}}$ for each case.

First, when we use the non-interacting Green function (i.e., the $\Sigma=0$ case), the FLEX pairing interaction yields a larger $\lambda_{\mathrm{SC}}$ than the D$\Gamma$A pairing interaction, except for $n=0.95$, where spin fluctuations are strongly developed. This indicates that the pairing interaction in FLEX is stronger than in D$\Gamma$A. 
The smaller D$\Gamma$A values away from $n=0.95$ may reflect the reduction of the low-frequency pairing interaction by vertex corrections, as discussed in Ref. [\onlinecite{Kitatani2019}].
On the other hand, for a fixed D$\Gamma$A pairing interaction in Fig.~\ref{fig:modified-self-energy}(a), using the FLEX self-energy results in a reduction of $\lambda_{\mathrm{SC}}$.
This indicates that the suppressive effect of the FLEX self-energy on superconductivity is stronger than that of the D$\Gamma$A self-energy outside the pseudogap-like region [see also Fig.~\ref{fig:resultU8}(c,g)]. A similar trend can also be seen in previous comparisons for the single-layer Hubbard model \cite{Deng2026}.

Moreover, in the superconducting region, FLEX exhibits a pronounced difference between the $\Sigma=0$ and $\Im\Sigma=0$ cases. Here, $\Im\Sigma=0$ means that only the imaginary part of $\Sigma$ is set to zero, while the real part is kept unchanged. The large difference between these two cases reflects the strong $\Re\Sigma$-induced reconstruction of the low-energy spectrum in FLEX, including the increase in the effective bonding-antibonding splitting discussed above.

In contrast, D$\Gamma$A also shows a sizable difference between the $\Sigma=0$ and $\Im\Sigma=0$ cases particularly at $n=0.95$, but its origin appears to be different from that in FLEX. As shown in Fig.~2 of the main text, D$\Gamma$A does not show the pronounced $\Re\Sigma$-driven enhancement of the bonding and antibonding band splitting found in FLEX. The absence of such a pronounced splitting enhancement is consistent with the previous cellular DMFT result \cite{Nomura2025}. 
\\

\bibliography{main}